\documentclass[runningheads]{llncs}
\usepackage{amssymb,amsmath}
\usepackage{pstricks}
\usepackage{pst-node}
\usepackage{epsfig}
\usepackage{latexsym}
\begin{document}
\pagestyle{headings}

\newcommand{\MUB}{{\sf mub}}
\newcommand{\mub}{\,\Join\!}
\newcommand{\cx}{{\sf Cx}}
\newcommand{\mubdown}[2]{\widehat{\Join}^{#1}\!(#2)}
\newcommand{\mupdown}[2]{\widetilde{\Join}^{#1}\!(#2)}
\newcommand{\Id}{{\sf Id}}
\newcommand{\covers}{\succ}
\newcommand\QED{\vskip 0pt $\hfill\Box$ \vskip 1pt}
\newcommand{\fps}[1]{\!<\!\!<\! #1\! >\!\!>}
\newcommand{\length}[1]{\mid\! #1 \!\mid} 
\newcommand{\covered}[1]{\lfloor \! #1 \! \rfloor}
\newcommand{\cover}[1]{\lceil \! #1 \! \rceil}
\newcommand{\Con}{\textit{Con}}
\newcommand{\elements}[1]{|#1|}
\newcommand{\default}[3]{\dfrac{#1 : #2}{#3}}
\newcommand{\nmc}{\ensuremath{\mid\!\sim}}

\title{Axiomatic Aspects of Default Inference
\thanks{Originally published in proc. PCL 2002, a FLoC workshop;
eds. Hendrik Decker, Dina Goldin, J{\o}rgen Villadsen, Toshiharu Waragai
({\tt http://floc02.diku.dk/PCL/}).}
}
\author{Guo-Qiang Zhang}

\institute{Department of EECS\\
Case Western Reserve University\\
Cleveland, Ohio 44022, U.S.A.\\
\email{http://newton.cwru.edu/$\!\sim$gqz\\
gqz@eecs.cwru.edu}}  

\maketitle

\begin{abstract}
Properties of classical (logical) entailment relation
(denoted as $\vdash$) have been well studied and well-understood,
either with or without the presence of logical connectives.

There is, however,  less uniform agreement on laws for the nonmonotonic
consequence relation.
This paper studies axioms for nonmonotonic consequences
from a semantics-based point of view, focusing on a class of mathematical
structures for reasoning about
partial information without a predefined syntax/logic. This structure is
called a default structure.
We study axioms for the nonmonotonic consequence
relation derived from  {\em extensions} as in Reiter's default logic,
using {\em skeptical  reasoning}, but extensions are now used 
for the construction of {\em possible worlds} in a default information structure.

In previous work we showed that skeptical reasoning arising from 
default-extensions obeys a well-behaved set of axioms including
the axiom of cautious cut. We show here that, remarkably, the
converse is also true:
{\em any} consequence relation obeying this set of axioms can be
represented as one constructed from skeptical reasoning.
We provide representation theorems
to relate axioms for nonmonotonic
consequence relation and properties about extensions,  and provide
a {\em one-to-one correspondence} between nonmonotonic systems which
satisfies the law of {\em cautious monotony} and default structures with
{\em unique} extensions. Our results give a  theoretical justification for
a set of basic rules governing the update of nonmonotonic knowledge bases,
demonstrating the derivation of them  from the more concrete and primitive
construction of extensions. It is also striking to note that  proofs of the representation
theorems show that only  {\em shallow extensions} are necessary,
in the sense that the number of iterations needed to achieve an extension is at most three.
All of these developments are made possible by taking a more liberal view of
consistency: {\em consistency is a user defined predicate},
satisfying some basic properties. \end{abstract}

\bigskip

\section*{Introduction}

{\em Reasoning} in general is concerned with drawing conclusions
from a set of premises. Mathematical reasoning is monotonic,
in the sense that if the set of premises becomes bigger, 
the set of conclusions also grows larger (or stays the same). Reasoning in daily life,
however, is nonmonotonic, because conclusions drawn earlier
due to the lack of information may have to be withdrawn later
in light of new information. This nonmonotonic phenomenon can be
avoided by putting time-stamps on conclusions. Thus, the
conclusions drawn earlier and the conclusions withdrawn later 
have different time-stamps and are treated as different conclusions.
The cost of this temporal approach, which will not be discussed 
further here, is the 
``frame problem'', referring to the issue of keeping track of
much of background information that stays unchanged 
as time goes on.  

Nonmonotonicity arises when explicit time-stamps are avoided when
we are in fact dealing with time-sensitive, partial information.
For example, in databases data are not 
time-stamped, but they evolve continuously over time. 

A well-known formalism for nonmonotonic reasoning is 
Reiter's default logic \cite{reiter}, in which the so-called
{\em extension} construction was introduced as a 
important  method to extend current knowledge
using default rules. Reiter's treatment is {\em syntactic}, in the sense
that default rules are used as extended {\em proof rules} for first order logic,
allowing more conclusions derived provided that
the global property of  {\em consistency} is not violated. It should be of no surprise 
that this treatment has caused some anomalies, because
first order logic was invented for {\em mathematical} reasoning rather than
commonsense reasoning.

There is  a conceptually cleaner view of default reasoning 
under the paradigm of modal logic with Kripke structure as models \cite{huang}.
We can think of a state in a Kripke structure as 
a state of knowledge/information. Possible worlds represent 
hypothetical worlds which may evolve from a given state.
Modal logic can then be adapted for the purpose  of
reasoning about belief. The role of default is now relegated
to the semantic structure as a concrete way for
{\em constructing} possible worlds using extensions.
Nonmonotonic reasoning in this new setting becomes
{\em model-checking}: given a state (a knowledge base) in  
the Kripke structure and
a belief formula (expressed in modal logic), determine
whether or not  the belief is supported. Nonmonotonicity corresponds to
the fact that a belief supported in the current state may become
unsupported in a possible future state. 

This semantic view of default reasoning is developed in detail
in the so-called default domain theory \cite{rounds}. The idea is to use Scott's
{\em information system} \cite{scott} as the basic semantic structure for representing
information states (the underlying Kripke structure), extended with default rules in order
to construct possible worlds. Here are
some specific achievements of default domain theory.

\begin{itemize}
\item {\em Power default reasoning.} Domain theory is a powerful and elegant
framework developed by Scott and others for the denotational semantics of programming
languages. A routine practice in domain theory is that 
an object of one type can be embedded/projected to different type in order to facilitate a more
appropriate treatment. The  idea of power default reasoning \cite{zw3} is to 
encode default rules in a higher-order setting
so that a {\em nonmonotonic operator at the base level} induces a better behaved
{\em operator in the higher-order space of Smyth powerdomain}.
Under power default reasoning,  the  extension construction has nice
structural properties.
For example,  one can show a dichotomy theorem \cite{zw3} which states that
with respect to a set of default constraints  on a Scott domain $D$ lifted as power defaults  $\Gamma$
on the Smyth powerdomain ${\cal P}(D)$,  an element either has a safe, unique
 $\Gamma$-extension, or else  the multiple  $\Gamma$-extensions will all be singleton generated.
The Extension Splitting Theorem \cite{zw3} states that any extension of
the union of  two compact open  sets can be split into the union of  two corresponding extensions.
This allowed us
to prove, among other things,  the law of  reasoning by cases and  the  law of cautious
monotony, as discussed in \cite{klm}. 
Note that the cases law and cautious
monotony law  do not hold  for standard propositional default 
logic, which is among the anomalies of default logic discussed in the literature.

\item {\em Complexity.} These structural properties have direct algorithmic consequences.
Based on the  Dichotomy Theorem,  an
algorithm \cite{zw2} has been developed for  skeptical normal default inference in propositional 
logic to show that the problem is complete for co-NP(3), the third level of the {\em 
Boolean hierarchy}.  
This contrasts favorably with standard propositional default reasoning
based on default logic, which was proved to be
$\Pi^{\rm P}_{2}$-complete \cite{gottlob,eiter}.

\item {\em Semantics of disjunctive logic programming.}
Our domain-theoretic investigation to logic programming started with the
basic observation that the {\em information system} representation of
domains \cite{scott} bears a remarkable similarity to the syntax of definite logic
programs (the so-called Horn clauses). Our work \cite{rz} shows that
a general disjunctive logic programming paradigm can be developed
on coherent domains -- algebraic cpos on which the intersection of any two
compact open sets remains compact open. From the domain-theoretic point of
view, these are very general spaces which contain several prominent 
categories already (such as Scott domains and SFP domains).
More concretely,  a disjunctive logic program can be regarded as
a {\em sequent structure}. Such a structure generates  {\em spatial locales}
 (the so-called pointless topology \cite{johnstone}),
which provide models and proof rules for which completeness is guaranteed \cite{coquand}.

\end{itemize}

\bigskip
In this paper we continue the semantics-based study of nonmonotonic inference by
providing representation theorems for the skeptical nonmonotonic consequence relation.
We show that an abstract nonmonotonic relation satisfies the law of cautious monotony
(among other reasonable axioms) if and only if it can be generated from a set of 
normal default rules {\em with unique extensions}.
We then provide some preliminary discussions on a possible  categorical setting in which to discuss properties of these
constructions.
These developments are made possible by taking a less restricted view of
consistency: we view {\em consistency as a user defined predicate},
satisfying some basic properties, as precisely captured
by {\em information systems} -- $\Con$.
There is no prescribed global notion of consistency;
a set of tokens can be consistent in one information system while
inconsistent in another. 
We believe that the meaning for  ``paraconsistency'' ultimately lies in this local, user-defined
interpretation.
\bigskip

{\bf Related work}. The idea of extension was introduced by Reiter \cite{reiter}.
Makinson \cite{makinson} and others \cite{garden}
have provided extensive study of many axioms discussed here, sometimes
from the belief-revision point of view. Marek, Nerode, and Remmel \cite{mnr} 
have studied structures  similar to default information structures,
from a recursion-theoretic point of view.
The first representation result similar to the ones 
studied here was  given in \cite{zw1}.
Hitzler and Seda \cite{seda,sehi}, in a sequence of papers, studied logic programming 
semantics using the tools of Topology and Analysis. The edited volume \cite{tan}
contains a number of discussions on the role of partiality of information for
commonsense reasoning. These are philosophically
allied to our domain-theoretic approach to logic programming and nonmonotonic reasoning.
 Fitting \cite{fitting,fittingb,fittingc} was among the first ones to introduce order to the study 
of logic programming in a systematic way. There are also other kinds of representation results
in the literature. For example,  Marek, Treur, and Truszczynski \cite{marek}
studied the problem of when a family of theories can be represented as the extension family of
normal defaults. Our work differs from this in that we relate default theory to {\em axioms for
the nonmonotonic consequence relation},  establishing connections between two
independent paradigms.

\section{Default domain theory}

This section provides an overview of the
basic structure of default domain theory -- default information structures.
We  refer to \cite{scott,zh} for  more discussion on information systems, and
\cite{besnard,brewka,mtr}
for background on nonmonotonic reasoning.

\subsection{Information systems}

In order to study the properties of nonmonotonic systems, 
let's first take a look at monotonic systems, as captured in  
Scott's {\em information systems}.

An information system consists of a set $A$ of tokens, a subset 
$\Con$ of the set of finite subsets of $A$, denoted as ${\rm Fin} (A)$, 
and a relation $\vdash$ between $\Con$ and $A$. The subset $\Con$ on 
$A$ is often called the consistency predicate, and
the relation $\vdash$ is called the entailment relation.
Both the consistency predicate and the entailment relation
satisfy some axioms due to Tarski, made precise in the following definition.

\begin{definition} 
An information system $\underline{A}$ is a triple $( A, \Con, \vdash),$
 where
\begin{enumerate}
\item $A$ is the token set,
\item $\Con$ is  the consistency predicate ($\Con \subseteq {\rm Fin}  
(A)$ and $\emptyset \in \Con$),
\item $\vdash$ is the entailment  relation ($\vdash \; \subseteq \;
 \Con\times A$).
\end{enumerate}
Moreover, the consistency predicate and  entailment  relation  satisfy
the following properties:
\begin{enumerate}
\item  $X\subseteq Y\;{\&}\; Y\in \Con \Rightarrow X\in \Con ,$
\item $a\in A\Rightarrow \{\, a \,\}\in \Con ,$
\item $X\vdash a \; {\&}\; X\in \Con \Rightarrow X\cup \{\, a\, \}\in \Con ,$
\item $a\in X\;{\& }\; X\in  C\!on\Rightarrow X\vdash a ,$
\item $ ( \forall b\in Y\, X\vdash b) 
\;{\&}\; Y\vdash c \Rightarrow X\vdash c .$
\end{enumerate}
\end{definition} 

Although monotonicity for $\vdash$ is not explicitly given, it is a derivable
property. Suppose $Y\vdash a$ and $Y\subseteq X$ with $X\in \Con$.
By $(4)$, $X\vdash b$ for every $b\in Y$. Now apply $(5)$ and we get
$X\vdash a$. Thus monotonicity is an inherent property of $\vdash$:
$$Y\vdash a\;\&\; Y\subseteq X \Rightarrow X\vdash a.
\footnote{Here we deliberately dropped the $X\in\Con$ condition.
One can, as in logic, define $\Con$ in such a way that
$X\in \Con$ if and only if $X\not\vdash a$ for some $a\in A$, as 
suggested by Peter Aczel in a personal communication. However, this ``logical''
definition is more limited, in that it insists on the whole set $A$ to be inconsistent.}$$

The notion of consistency can be easily extended to arbitrary token sets 
by enforcing  compactness, i.e., a set is consistent if every 
finite subset of it is consistent. By overloading notation, we 
write $X\in \Con$ when every finite subset of $X$ is consistent. 

The monotonicity of $\vdash$ induces a monotonic
operator $F: \Con \to \Con $ for an information system $( A, \Con, 
\vdash)$:
$F(X) : = \{ a \mid \exists Y ( Y\subseteq^{\rm fin} X \;\&\;
Y\vdash a)\}.$
(Here, $\subseteq^{\rm fin}$ stands for ``finite subset of''.)
One can show, by property $(3)$ of $\vdash$, that $F(X)$ is consistent if $X$ is 
(so $F$ is a well-defined function).

The reflexivity property $(4)$ implies that $F$ is {\em inflationary},
in the sense that $X\subseteq F(X)$ for any $X\subseteq A$.
The continuity of $F$ follows from the finiteness (or compactness)
of $\vdash$ on its left-hand-side.

From these we can easily show that for any given subset $X$ of
tokens, $F$ has a least fixed-point containing $X$.
 $F$ is in fact a
{\em closure} operator (often denoted as $C\!n$ in the literature), 
with the following defining properties:
\begin{itemize}
\item Inflationary: $X\subseteq F(X)$;
\item Monotone: $X\subseteq Y\Rightarrow F(X)\subseteq F(Y)$;
\item Idempotent: $F(F(X)) = F(X)$.
\end{itemize}
 
Idempotency of $F$ follows from the transitivity $(5)$ of
$\vdash$. To show the non-trivial containment $F(F(X)) \subseteq  F(X)$,
let $a\in F(F(X))$. This means that for some finite $Y\subseteq F(X)$,
$Y\vdash a$.  But for each $b\in Y$, there is some 
finite $X_{b}\subseteq X$
such that $X_{b}\vdash b$. In fact, the finite subset
$\bigcup \{X_{b}\mid b\in Y\}$ of $X$ entails every token in $Y$, by
monotonicity. Now the transitivity of $\vdash$ gives
$\bigcup \{X_{b}\mid b\in Y\}\vdash a$ and so $a\in F(X)$, as required.

The {\em information states}  of an information system are precisely
sets of the form $F(X)$ with $X\in \Con$, where $\Con$ is 
understood in the generalized sense to include infinite sets through the
compactness condition.
Moreover, the set of all information state
 $\{ F(X) \mid X\in \Con\}$ is precisely the
set of fixed-points of $F$.

The importance of information systems lies in the fact that they
provide a concrete representation of Scott domains\footnote{Exactly what
a Scott domain  is need not concern us here. It should suffice to say that
a Scott domain is a complete lattice with the top element removed.}.

\begin{theorem} (Scott) For any information system 
$\underline{A}$, the collection 
of its information states $\elements{\underline{A}}$
under inclusion forms a Scott domain. Conversely, every 
Scott domain is order-isomorphic to the partial order of information states
 of some information system.
\end{theorem}

\subsection{Normal default structures}

We now introduce the main definitions of default domain theory. Normal 
default structures are information systems extended with a set of 
default rules of the form 
$\default{X}{a}{a},$
with $X$ a finite consistent set, and $a$ a single token of the underlying 
information system.
The idea of a default rule is that one can generate models (states of belief) by 
finding $X$
as a subset of tokens in a state under construction, checking that $a$ is 
consistent with the current state, and then adding the token $a$. 

\begin{definition}
A normal default  structure is a tuple 
$ \underline{A}=(A,\,\Con , \Delta ,\,\vdash )$
where $(A, \Con ,\vdash )$ 
is an information system,
$\Delta$ is a set of normal defaults,
each element of which is written as
$\default{X}{a}{a}$, with $X\in\Con$, $a\in A$. If each default is of the 
form $\default{\emptyset}{a}{a}$, we call the default structure precondition free. 
\end{definition}

Extensions are a key notion related to a default structure. An extension
of an information state
 $x$ is intuitively an information state $y$ extending $x$, constructed in 
such a way that everything in $y$ reflects an agent's belief 
expressed by defaults. If the current situation is $x$, then because it is 
a partial model, it may not contain enough information to settle an issue 
(either positively or negatively). Extensions of $x$ are partial models 
containing at least as much information as $x$, but the extra information 
in an extension is only plausible, not factual.

The following definition is just a reformulation, in information-theoretic 
terms, of Reiter's own notion of extension in default logic. 

\begin{definition}\label{extension-def}
Let $\underline{A}=(A,\Con , \Delta ,\vdash )$ be a default
structure, and $x$ an information state of the information system
$(A,\Con , \vdash )$. For any
$S\subseteq A$,
define $\Phi (x, S)$ to be the union $\bigcup_{i\geq 0} \phi (x, S, 
i)$, where (note that $F$ is given in the previous subsection)
\[
\begin{array}{l}
\phi (x, S, 0)=x,\\[1ex]
\phi (x, S, i+1)=
\begin{array}[t]{l}
F(\phi (x, S, i) ) ~ \cup \\[1ex]
\{ a \mid 
\default{X}{a}{a}\in \Delta \;{\&}\; X\subseteq \phi (x, S, i)\;{\&}\; 
\{a\}\cup S\in C\! on \}.
\end{array}
\end{array}
\]
Call $y$ an extension of $x$ if $\Phi (x, y)=y$. In this case we also 
write $x\, \epsilon_{\underline{A}} \, y$, with the subscript omitted when
the default structure under consideration is clear. \end{definition}

Basic properties of extensions are stated next, which will be used
in the proof of main  representation theorems in the next section. 
These properties are parallel to those for Reiter's
default logic, and so their proofs are omitted.

\begin{theorem}\label{existence}
Let  $\underline{A}=(A,\Con , \Delta ,\vdash )$ be a normal default  
structure, and $x, y$ information states of the information system
$(A,\Con , \vdash )$.
We have
\begin{enumerate}
\item $x$ has at least one extension.
\item if $x\, \epsilon\, y$ then $y\supseteq x$. 
\item if $x\, \epsilon\,  y$, then
$y\, \epsilon \, z$ if and only if $y=z$ for any information state $z$.
\item for any information state $z$,
if $x\, \epsilon\,  y$ and $ x\, \epsilon\, z$, then either $y=z$ or $y\cup z\not\in \Con$. 
\end{enumerate}
\end{theorem}

In terms of possible world semantics, normal  default structures give
rise to Kripke structures in which every state has either a transition to
a different state or loops to itself. Moreover, if a state does contain a transition
to a new state, then such future states are pairwise incompatible
(there is no common future states).

\section{Axioms for nonmonotonic consequence} 

There is little disagreement about what properties should
a (monotonic) entailment relation have: these were
given by Tarski, as captured in information systems.
The situation is quite different for
the nonmonotonic consequence relation.
To motivate the discussion, we first take a look at
the {\em skeptical nonmonotonic consequence relation}
determined by a normal default information structure
{\em with a trivial $\vdash$}.
The default structures will have only three components, as
$(A,\Con ,  \Delta )$, with  $\vdash$ understood as
$X\vdash a$ if and only if $a\in X$.
The closure operator $F$ will now be the identity function:
$F(X)=X$.

\begin{definition}\label{nmc}
The skeptical nonmonotonic consequence relation $\nmc_A$
with respect to a normal default structure $(A,\Con ,  \Delta )$ 
 is defined as $X\nmc _A a$
if $a$ belongs to every extension $y$ of $X$.
Here $X$ is a finite consistent subset of $A$, and $a\in A$. 
\end{definition}

Since  extensions exist in normal default 
structures, we have
$$X\nmc_{A} a\Longleftrightarrow a\in \bigcap \{ y\mid 
X\epsilon_{A} y\} .$$

The next example shows that $\nmc_{A}$ is not monotonic:
$X\nmc_{A}a$ and $X\subseteq Y$  does not imply $Y\nmc a$.

\begin{example}
Consider the default structure $(A,\Con ,  \Delta )$ with
$A:=\{ a, b\}$,
$\Delta :=
\{ \default{\emptyset }{b}{b}\}$, and 
$\{ a, b\}\not\in \Con$.

There is a unique extension for $\emptyset$: $\{b\}$. There is only
one extension for $\{a\}$ as well:
$\{a\}$ itself. The conflict with $b$ prevents us from adding $b$
to $\{a\}$.
We have $\emptyset\nmc b$,
but $\{ a\}\not \nmc b$.
\end{example}

Failing monotonicity, what other properties can we say about
$\nmc_{A}$? For sake of brevity, we write, for finite sets
$X$ and $Y$, $X\nmc Y$ to mean that
$X\nmc b$ for every $b\in Y$.
Certainly we have $X\nmc X$ for every consistent finite set $X$.
The only non-trivial property we can say about $\nmc$ 
in general seems to be {\em cautious cut}:
$X\nmc T\;{\&}\; T , X\nmc Y\Rightarrow X\nmc Y.$
We call it cautious cut because the standard cut axiom takes the 
following form:
$$ X\nmc T\;{\&}\; T, Y\nmc Z\Rightarrow X, Y\nmc Z.$$ 
This is equivalent to cautious cut when monotonicity is assumed, 
but cut is stronger in the nonmonotonic case.

We need the following property to prove cautious cut.

\begin{lemma}\label{cut}
For any finite consistent sets $P, Q$ of a default structure
$(A,\Con, \Delta )$, 
if $P\, \epsilon\,  R$
and $Q\, \subseteq \, R$, then $(P\cup Q)\, \epsilon\, R.$ 
\end{lemma}

A  proof for this can be found in \cite{zw1}.

\begin{theorem}\label{cautious-cut}
Let $(A, \Con ,\Delta )$ be a normal default structure.
The derived skeptical nonmonotonic relation $\nmc_{A}$
satisfies cautious cut.
 \end{theorem} 

\begin{proof}  Let $X\nmc _A T$ and 
$ T, X\nmc _A Y$ for finite consistent sets $X, T, Y$.
We have
$T\subseteq \bigcap \{ e\mid X\epsilon_A e\}$ and 
$Y\subseteq 
\bigcap \{ e\mid (X\cup T)\epsilon_A e\}.$

We need to show that $Y\subseteq 
\bigcap \{ e\mid X\epsilon_A e\}.$
 Let $e$ be an 
extension of $X$. We have $T\subseteq e$ since $X\nmc _A T$. 
By Lemma~\ref{cut}, $e$ is an 
extension of
$X\cup T$. However, $Y$ is a subset of every extension of 
$X\cup T$; in particular,
$Y\subseteq e$.
  Therefore, $Y$  is a subset of every extension of 
$X$, as required for $X\nmc_{A}Y$.\QED
\end{proof}

We take the properties satisfied by  $\nmc_{A}$
as the  minimal set of properties that any
nonmonotonic consequence relation should satisfy.
This brings us to the notion of {\em abstract nonmonotonic system}.

\begin{definition}\label{abstract}
An abstract nonmonotonic system is a triple
$(A, \Con , \nmc )$,
where $\Con$ is a collection of finite subsets $X$ of $A$, called the 
consistent sets,
$\nmc $ is a subset of $ \Con \times \Con$, 
called the relation of nonmonotonic entailment, 
which satisfies the  following axioms:
\[\begin{array}{l}
1.\; X\subseteq Y\in \Con \Rightarrow X\in \Con ,\\ 
2.\; a\in A\Rightarrow 
\{a\}\in \Con ,\\ 
3.\; X \nmc T\Rightarrow X\cup T\in \Con ,\\ 
4.\; Y\subseteq X\Rightarrow X\nmc Y, \\ 
5.\; X\nmc T\;{\&}\; T, 
X\nmc Y\Rightarrow X\nmc Y,\\ 
6.\; X\nmc Y\;{\&}\; X\nmc Z\Rightarrow X\nmc Y\cup Z.\end{array}\]
\end{definition}

Note that Axiom 4 is reflexivity, and Axiom 5 is cautious cut.
The rest of the properties are routine.
Axioms 4 and 6 together allow us to view a
 nonmonotonic system as one generated from instances of the
 form $X\nmc \{a\}$. One can then 
 define $X\nmc Y$ if and only if $X\nmc \{b\}$ for every
 $b\in Y$.   $X\nmc \emptyset$ is vacuously true.

\begin{theorem}[Zhang and Rounds \cite{zw1}]\label{rep-half}
For any  a normal default structure $(A, \Con, \Delta )$,
$(A, \Con , \nmc_{A})$ is an abstract nonmonotonic
system, where the relation $\nmc_{A}$ is given in Definition~\ref{nmc}.
\end{theorem}

We will not go through the formality of the proof here, but
only note that cautious cut follows from Theorem~\ref{cautious-cut},
and Axiom 3 follows from the fact that extensions always exist
and each one of them is a consistent set.

\section{Representation theorems}

Theorem~\ref{rep-half} says that any normal default structure
determines an abstract nonmonotonic system -- a nonmonotonic
relation satisfying a minimal set of properties including reflexivity 
and cautious cut.

The more interesting question is the converse:
is it true that {\em every} abstract nonmonotonic system
can be represented concretely as one derived from a normal
default structure? In other words, for each abstract nonmonotonic system
$(A, \Con , \nmc)$, is there a normal default structure
 $(B, \Con, \Delta )$ such that $\nmc \; = \;\nmc_{B}$?

We show that the converse of Theorem~\ref{rep-half} is indeed true.
Such a representation result bears much resemblance to 
the fundamental Cayley's Theorem  for finite groups, saying that
any finite group is isomorphic to a permutation group.
The next theorem says that
every abstract nonmonotonic relation is determined by 
some normal default structure.
This shows that {\em the framework of default rules with extensions,
 although a concrete nonmonotonic formalism, is expressive
 enough to represent any reasonable nonmonotonic consequence relation}.

\begin{theorem}\label{rep-onlyif}
Let $(A, \Con , \nmc )$ be an abstract nonmonotonic system. There is a normal 
default structure $\underline{B}=(B, \Con^{*},\Delta )$ satisfying
the following properties:
\begin{enumerate} 
\item $B\supseteq A$, 
\item for any $X\subseteq B$,
 $\;\;X\in \Con \;\;\hbox{\rm if and only if } \; X\subseteq A\;\hbox{\rm and } X\in \Con^{*},$
\item for every $X, Y\in \Con$,
$\;\; X\nmc Y \;\; \hbox{ if and only if  }\; X\nmc_{B} Y.$
\end{enumerate}
\end{theorem}

Conditions 2 states that $\Con^{*}$ is a conservative extension of $\Con$, and
condition 3 says that $\nmc_{B}$ is a conservative extension of $\nmc$.

\begin{example}
An example will be helpful to illustrate the idea of the proof. 
Consider the abstract  nonmonotonic system $(A, \Con , \nmc )$
with $A=\{a, b\}$, $\Con = 2^{A}$ and
$\nmc$  generated by requiring $\emptyset\nmc \{a\}$ and 
$\emptyset\nmc \{b\} $ (i.e., reflexivity and cautious cut is always 
assumed; note that we have neither 
$\{a\}\nmc \{b\}$, nor  $\{b\}\nmc \{a\}$).
We would 
like to construct a default structure $\underline{B}=(B, \Con^{*}, 
\Delta)$ which determines 
this nonmonotonic entailment relation\footnote{This example also explains
why we cannot in general require $B=A$ in Theorem~\ref{rep-onlyif}. Here, if we
only used tokens $a$ and $b$, then for any default structure giving rise to $\emptyset\nmc \{a\}$ and 
$\emptyset\nmc \{b\} $, we would also have had $\{a\}\nmc \{b\}$ -- try it!}.

The idea is to introduce, for each consistent set $X$,
a new token $[X]$, similar to the {\em powerset construction}
in automata theory. 
Thus we have a total of 4 new tokens:
$[\emptyset], 
[\{a\}], [\{b\}],
[\{a,b\}].$

For each new token $[X]$, we introduce a default
rule $\default{X}{[X]}{[X]}$ as well as a set of 
default rules
$\{ \default{\{ [X] \}}{a}{a} \mid  X\nmc a\;\&\; a\not\in X \}.$
For the example at hand, we have 
the following default rules:
\[\begin{array}{l}\default{\emptyset}{[\emptyset]}{[\emptyset]}, 
\default{\{[\emptyset]\}}{a}{a},
\default{\{ [\emptyset]\}}{b}{b},\\
\default{\{a\}}{[\{a\}]}{[\{a\}]},
\default{\{b\}}{[\{b\}]}{[\{b\}]},
\default{\{a,b\}}{[\{a,b\}]}{[\{a,b\}]}.\\
\end{array}\]

The consistency predicate $\Con^{*}$
 is defined in such a way that
it extends $\Con$, but  new tokens are inconsistent with each other.
Moreover, for a set containing a new token
$[X]$ to be consistent, every other element in the set
must be a nonmonotonic consequence of $X$.
Therefore, the consistent sets in $\Con^{*}$ are
sets (as well as their subsets) of the form
$Y\cup \{ [X] \}$
with $X, Y\subseteq \{a, b\}$,
$X\nmc Y$.
For instance, 
$\{[\emptyset], [\{a\}]\}$ and 
$\{[\{a\}], b\}$ are inconsistent sets.

We have
$\emptyset \nmc _B \{a\}$ and $\emptyset \nmc _B \{b\}$,
because the unique extension for $\emptyset$ is
$$\{ [\emptyset], a, b\}.$$
But we do not have $\{b\} \nmc _B \{a\}$,
 since there are two extensions for $\{b\}$:
 $$\{b, [\{b\}]\}\;\;\hbox{\rm and }\; \{a, b, [\emptyset ]\}, $$
 and $a$ is not in the first extension.
Similarly, we do not have
$\{a\}\nmc _B \{b\}$ either.
\end{example}

\bigskip

We now describe a general procedure to construct the required default 
structure $\underline{B}=(B, \Con^{*} , \Delta )$ from an
abstract nonmonotonic system
$(A, \Con , \nmc )$.

The token set $B$ is $A\cup \{ [X] \mid X\in \Con \}.$
 The idea is to introduce a 
new token for each consistent set $X$ and use sets of given tokens to encode new tokens.
Note that we need only introduce $[X]$ for (consistent) finite sets $X$ in the powerset of $A$ here, even though $A$ may be infinite.

We have two kinds of  default rules.
One is $\default{X}{[X]}{[X]}$ for each $X\in \Con$, and the other
is $\default{\{ [X] \}}{a}{a} $ for each
nonmonotonic instance  $X\nmc a$ with $a\not\in X$. 
Thus
$$\Delta := \{ \default{X}{[X]}{[X]} \mid X\in \Con \}
\cup \{  \default{\{ [X] \}}{a}{a} \mid  X\nmc a\;\&\; a\not\in X \}.$$

For any subset $W$ of $B$, $W\in \Con^{*}$ if and only
if all of the following three conditions hold: 
\begin{enumerate}
\item $W\cap A\in \Con$, i.e., the old tokens in $W$ form a consistent
set in $\Con$;
\item at most one token of the form $[X]$ is in $W$;
\item if $[X]\in W$, then $X\nmc W\setminus \{[X]\}$.
\end{enumerate}

 It is straightforward to check that
 the consistency predicate defined  this way has the required 
properties. Each individual token is indeed consistent.
For any  subset $Z$ of $W\in \Con^{*}$, we have $Z\in \Con^{*}$
by examining the three conditions above.

The remaining task is to show that the
default structure $\underline{B}$ has the properties
as stated in Theorem~\ref{rep-onlyif}.
This will be achieved in several lemmas.

An auxiliary notation is needed.
We define 
$\widetilde{X}:= \{ t\mid X\nmc \{t\}\},$
i.e., $\widetilde{X}$ is the set of all nonmonotonic consequences
of $X$.

\begin{lemma}\label{nonmono1}
Let $(A, \Con , \nmc )$ be an abstract nonmonotonic system. We have, for any 
consistent set $X$,
$$\bigcap\{ \widetilde{Y}
\mid Y\subseteq X\subseteq \widetilde{Y}\}=\widetilde{X}.$$ \end{lemma}

\begin{proof} We prove the non-trivial direction
$\supseteq$ by showing that if $Y\subseteq X\subseteq \widetilde{Y}$
then $\widetilde{X}\subseteq\widetilde{Y}.$
Suppose $Y\subseteq X\subseteq \widetilde{Y}$. Given any $a$
in $\widetilde{X}$, we can rewrite this as 
$X\setminus Y, Y\nmc a$ since $Y$ is a subset of $X$.
 Further more, since
$X\subseteq \widetilde{Y}$, we have $Y\nmc X\setminus Y.$ Now, applying 
cautious cut we get $Y\nmc \{a\}$. Therefore 
$a\in \widetilde{Y}.$ \QED
\end{proof}

The axiomatization of the
 nonmonotonic operator $\widetilde{(\;\; )}$   has been studied
 in depth, notably by Makinson and others. Here we only indicate 
 that it is not monotonic (of course), and it is not necessarily true
 that $\widetilde{\widetilde{X}}=\widetilde{X}.$

\begin{lemma}\label{nonmono3}
Let $P, Q$ be finite consistent sets in an abstract nonmonotonic system
 $(A, \Con , \nmc )$ and $\underline{B}=(B, \Con^{*}, \Delta)$ the
 corresponding default structure.  If
$Q\subseteq P\subseteq \widetilde{Q}$, then
$\widetilde{Q}\cup \{[Q]\}$ is an extension 
of $P$ in  $\underline{B}$.
\end{lemma}

\begin{proof} 
It is easy to see that the set $\widetilde{Q}\cup \{[Q]\}$ is
consistent with respect to $\Con^{*}$. Let's write $\rho (Q)$ for
$\widetilde{Q}\cup \{[Q]\}$.

 We need to verify the defining equality
$\rho (Q) = \bigcup_{i\geq 0}
 \phi (P, \rho (Q), i).$
By definition,
$$\phi (P, \rho (Q), 1)=
P \cup \{ a \mid 
\default{X}{a}{a}\in \Delta \;{\&}\; X\subseteq P\;{\&}\; 
\{a\}\cup \rho (Q)  \in \Con^{*} \}.$$
The only  applicable rules in $\Delta$ are of the form
$\default{X}{[X]}{[X]}$ with $X\subseteq P$; but
only  $\default{Q}{[Q]}{[Q]}$ will meet the consistency
requirement.
 Therefore, $\phi (P, \rho (Q), 1) = P\cup \{[Q]\}$.

The second iteration $\phi (P, \rho (Q), 2)$ is the set
$$(P\cup \{[Q]\})\cup 
 \{ a \mid 
\default{X}{a}{a}\in \Delta \;{\&}\; X\subseteq  (P\cup \{[Q]\})\;{\&}\; 
\{a\}\cup \rho (Q)  \in \Con^{*} \}.$$
This time all and only tokens in $\widetilde{Q}\setminus Q$
gets added to $\phi (P, \rho (Q), 1)$, since the applicable 
default rules are of the form  
$\default{\{[Q]\}}{b}{b}$ with $b\in \widetilde{Q}\setminus Q$.
Therefore, we obtain
$\phi (P, \rho (Q), 2)= (P\cup \{[Q]\})\cup \widetilde{Q} = \rho (Q),$
as needed.
\QED
\end{proof}

This proof conveys the additional information that
extensions such as $\rho (Q)$ can be attained in two iterations.
This offers an explanation of why default reasoning need not be hard
(see \cite{trusc} as well), because deep iterations for building extensions
may not be needed.

\begin{lemma}\label{nonmono4}
Let $P$ be a finite consistent set in an abstract nonmonotonic system
 $(A, \Con , \nmc )$ and $\underline{B}=(B, \Con^{*}, \Delta)$ the
 corresponding default structure.
Then every extension of $P$ in $\underline{B}$ is of the form
 $\widetilde{Q}\cup \{[Q]\}$,
 with 
$Q\subseteq P\subseteq \widetilde{Q}$. \end{lemma}

\begin{proof} 
Suppose $W$ is an extension of $P$.
We have $W\not = P$ since otherwise 
$\default{P}{[P]}{[P]}$  is applicable, 
forcing  $[P]$ to be added to $W$.

By design (of $\Delta$), the first token added to
$\phi (P, W, 1)$ must be  of the form $[Q]$, with
$Q\subseteq P$. So $\phi (P, W, 1) = P\cup \{[Q]\}$
for some $Q\subseteq P$. We show that $W= \widetilde{Q}\cup \{[Q]\}$.

The consistency constraint requires that
$P\subseteq \widetilde{Q}$.
It also requires that no other tokens of the form $[X]$ belong
to $W$.
So in the second iteration
one can (and must) apply all defaults of the form
$\default{\{[Q]\}}{b}{b}$ with
$b\in \widetilde{Q}\setminus Q$.
This gives
$$\phi (P, W, 2) =P\cup \{[Q]\}\cup (\widetilde{Q}\setminus Q).$$
  Since $Q\subseteq P$, we
get $W= \widetilde{Q}\cup \{[Q]\}$.
\QED
\end{proof}

\bigskip

\begin{proof}[Theorem~\ref{rep-onlyif}]
The first two items are obvious from the definition of $\underline{B}$.
We show the third item: for every $X, Y\in \Con$,
$X\nmc Y$ if and only if  $X\nmc_{B} Y.$

Let $X, Y$ be consistent subsets of $A$ such that
$X\nmc Y$. By Lemma~\ref{nonmono3} and Lemma~\ref{nonmono4},
extensions of $X$ in $\underline{B}$
 are precisely sets of the form
$ \widetilde{Q}\cup \{[Q]\}$ with $Q\subseteq X\subseteq \widetilde{Q}$.
For any such $Q$, we have, 
by Lemma ~\ref{nonmono1},  $\widetilde{X}\subseteq \widetilde{Q}$.
Therefore $Y\subseteq \widetilde{Q}$ and $Y$ is a subset of every
extension of $X$ in $\underline{B}$. This proves
$X\nmc_{B} Y.$

Suppose, on the other hand, $X\nmc_{B} Y$ for subsets $X, Y$ of $A$.
By Lemma~\ref{nonmono3} and Lemma~\ref{nonmono4},
$Y\subseteq \bigcap \{  \widetilde{Q}\cup \{[Q]\}\mid Q\subseteq X\subseteq 
\widetilde{Q}\}.$
Remembering that $Y\subseteq A$, we see that
$Y \subseteq \bigcap \{ \widetilde{Q} \mid Q\subseteq X\subseteq 
\widetilde{Q}\}.$
This means $Y\subseteq \widetilde{X}$, by  Lemma ~\ref{nonmono1}.
Therefore, $X\nmc Y$.\QED
\end{proof}

\section{Cautious monotony and unique extension}

The axiom of cautious monotony, considered by Gabbay,
Kraus, Lehmann, and Magidor and others \cite{gabbay,klm}, takes the form
$$X\nmc \{a\}\;\&\; X\nmc \{b\} \Rightarrow X\cup \{a\}\nmc \{b\}.$$
(We will drop the 
$\{\;\;\}$ for singletons and write
$X\nmc a\;\&\; X\nmc b \Rightarrow X, a\nmc b$ as well.)
This axiom intuitively says that 
{\em a nonmonotonic 
consequence can always be added as a premise without
affecting what was originally entailed.}
It is arguably one of the most fundamental assumptions for
a nonmonotonic knowledge base to work: if your knowledge
base $X$ can support several beliefs, say $a,b$, then
you should be able to adjoin these beliefs to enlarge the
knowledge base to $X\cup \{a,b\}$. Now, applying the reflexivity axiom
-- Axiom 4 of Definition~\ref{abstract}, we get
$X\cup\{a\} \nmc b$.

An easy induction shows that cautious monotony is
equivalent to the more general form
$X\nmc Y\;\&\; X\nmc Z \Rightarrow X, Y \nmc Z,$
with $X, Y, Z$ being finite (consistent) sets. 
If an abstract nonmonotonic system $(A, \Con, \Delta )$
satisfies cautious monotony, we will call it
a {\em cumulative nonmonotonic system}.

Here is an example showing that the skeptical 
nonmonotonic consequence relation derived from a
generic default structure need not satisfy cautious monotony.

\begin{example}
Consider the default structure $(A,\Con , \Delta )$ with
\[\begin{array}{l}
A=\{ a, b, c\},\\
\Delta =
\{ \default{\emptyset }{a}{a},
\default{\{a\}}{b}{b},\;
\default{\{b\}}{c}{c}\},\\
\{ a, b, c\}\not\in \Con .\\
\end{array}\]

There is a unique extension for $\emptyset$: $\{a, b\}$. There are, 
however,
two extensions for $\{b\}$:
$\{ a, b\}$ and
$\{ b, c\}$, both are maximal consistent sets.
We have, therefore,
$\emptyset\nmc b$, $\emptyset\nmc a$,
but $\{ b\}\not \nmc a$.
\end{example} 

Intuitively,  cautious monotony fails here because
the added premise (such as $b$ here) may trigger the
firing of a default rule (such as $\default{\{b\}}{c}{c}$ here)
not previously applicable, leading to
a different extension  (such as $\{b,c\}$)
which blocks a given default rule  (such as
$\default{\emptyset }{a}{a}$). This prevents an existing
consequence (such as $a$) from being present in all extensions
of the changed situation.

Cautious monotony, however, does hold when
extensions are unique. The reason is that in this case,
a nonmonotonic consequence can be added without
changing the (unique) extension.

\begin{theorem}[Zhang and Rounds \cite{zw1}]
Let $(A, \Con , \Delta )$ be a normal default structure
such that extensions are unique for any information state.
Then the derived nonmonotonic consequence relation $\nmc_{A}$
is cumulative, i.e., it satisfies cautious monotony.
\end{theorem}

In the rest of the section we show (which is new) that
the converse of the above is also true,
leading to another representation result.

\begin{theorem}\label{rep-cm}
For any cumulative nonmonotonic system $(A, \Con , \nmc )$,
there exists a normal 
default structure $\underline{B}=(B, \Con^{*},\Delta )$ with
the following properties:
\begin{itemize} 
\item $B\supseteq A$, 
\item for any $X\subseteq B$, we have
 $X\in \Con$ if and only if $X\subseteq A$ and $X\in \Con^{*}$,
\item for every $X, Y\in \Con$, it is the case that
$X\nmc Y$ if and only if  $X\nmc_{B} Y,$
\item there exists a unique extension for every information state of
 $\underline{B}$.
\end{itemize}
\end{theorem}

Since any cumulative nonmonotonic system  is an abstract
nonmonotonic system in the sense of Definition~\ref{abstract},
Theorem~\ref{rep-onlyif} tells us that it can be represented by a
normal default structure. The point of Theorem~\ref{rep-cm} is that,
we can require the normal default structure to have {\em unique extensions}, which
the construction given earlier for Theorem~\ref{rep-onlyif}
does not provide\footnote{Note that cautious monotony may still hold in the case of
multiple extensions. In fact, the
nonmonotonic consequence relations determined by
{\em precondition-free} default structures satisfy cautious monotony,
as pointed out in \cite{zw1}.}.

A lemma is needed to support the new construction.

\begin{lemma}\label{nonmono5}
Let $(A, \Con , \nmc )$ be a cumulative nonmonotonic system.
For finite consistent sets $P, Q$ of $A$, if
$Q\subseteq P\subseteq \widetilde{Q}$, then
$\widetilde{P} =\widetilde{Q}$.
\end{lemma}

\begin{proof}
Lemma~\ref{nonmono1} gives $\widetilde{P} \subseteq \widetilde{Q}$
and we need to show the other way round.

Suppose $Q\nmc a$. We need to show that
$P\nmc a$.
Since $P\subseteq \widetilde{Q}$, we have
$Q\nmc P$. Applying cautious monotony to
$Q\nmc P$ and $Q\nmc a$, we get
$P\cup Q\nmc a$. This is the same as $P\nmc a$ since $Q$
is a subset of $P$, by assumption.\QED
\end{proof}

One can see from the proof of Theorem~\ref{rep-onlyif}
that there is a 1-1 correspondence among sets $Q$s with
$Q\subseteq P\subseteq \widetilde{Q}$ and extensions of $P$,  
labeled by $[Q]$. The previous lemma suggests that
we should now collect all such $[Q]$s in one extension,
with the proviso that $Q\subseteq \widetilde{P}$ and
$\widetilde{P} =\widetilde{Q}$.

This is in fact the key idea for the proof of Theorem~\ref{rep-cm}.
Let $(A, \Con , \nmc )$ be an abstract nonmonotonic system
satisfying cautious monotony. Define 
$\underline{B}=(B, \Con^{*},\Delta ),$
where
\begin{itemize}
\item $B:= A\cup \{ [X] \mid X\in \Con\}$,
\item $\Delta :=  \{ \default{X}{[X]}{[X]} \mid X\in \Con \}
\cup \{ \default{\{[X]\}}{a}{a}\mid a\in \widetilde{X}\setminus X \}$,
\item $W\in \Con^{*}$ if and only
if all of the following three conditions hold: 
\begin{enumerate}
\item $W\cap A\in \Con$, i.e., the old tokens in $W$ form a consistent
set in $\Con$;
\item for each pair $[X], [Y]$ of (new) tokens in $W$,
$\widetilde{X}=\widetilde{Y}$;
\item if $[X]\in W$, then $X\nmc W\cap A$.
\end{enumerate}
\end{itemize}

Note that the difference from the previous construction is
in the consistency predicate $\Con^{*}$.
Two distinct tokens $[X], [Y]$ need not to be in conflict,
as long as $\widetilde{X}=\widetilde{Y}$.

  Theorem~\ref{rep-cm} is now an immediate consequence of
  the following lemma.

\begin{lemma}\label{nonmono6}
Let $\underline{B}=(B, \Con^{*}, \Delta)$ be the normal
default structure for
a cumulative nonmonotonic system
 $(A, \Con , \nmc )$ (as given in the  procedure above).
Every finite consistent set $P\subseteq A$ has a
unique extension, which is
$$\delta (P):=\widetilde{P} \cup \{ [Q]\mid
Q\subseteq \widetilde{P}\;\&\; \widetilde{Q} =\widetilde{P}\}.$$
\end{lemma}

\begin{proof}  
As a sanity check, we see that  $\delta (P)$ is indeed
consistent.
Suppose $W$ is an extension of $P$ in $\underline{B}$.
We show that $W=\delta (P)$.
By definition,  $W$ satisfies the
equality
$W = \bigcup_{i\geq 0} \phi (P, W, i).$
Clearly
$W\not = P$, since $\default{P}{[P]}{[P]}$ is one of the default
rules in $\Delta$.
But the first default rules applied must add tokens of the form
$[X]$, so $W$ contains a token of the form
$[Q]$, with $Q\subseteq P$. Consistency requires further
that $P\subseteq \widetilde{Q}$, since $P\subseteq W$.
By Lemma~\ref{nonmono5}, $Q\subseteq P$
and $P\subseteq \widetilde{Q}$ implies $\widetilde{Q} = \widetilde{P}$.

Since $[Q]\in W$, all tokens 
of the form 
$[R]$ with $\widetilde{R} = \widetilde{P}$ are consistent with $W$.
Therefore,  
$\phi (P, W, 1)= 
P \cup \{ [R] \mid 
R\subseteq P\;{\&}\; \widetilde{R} = \widetilde{P}\},$
which contains $[Q]$, in particular.

In the next iteration, 
all the default rules 
$\default{\{[R]\}}{c}{c}$ with 
$R\subseteq P$ and $\widetilde{R} = \widetilde{P}$ get
applied.
In particular, all default rules 
$\default{\{[P]\}}{a}{a}$ with $a\in \widetilde{P}\setminus P$ get
applied,
 with the net effect of adding all members
of $\widetilde{P}$ in $\phi (P, W, 2)$.
So,
$\phi (P, W, 2)= 
\widetilde{P} \cup \{ [R] \mid 
R\subseteq P\;{\&}\; \widetilde{R} = \widetilde{P}\}.$

Different from the case for Theorem~\ref{rep-onlyif},
we have not yet reached a fixed-point in the second iteration.
Additional default rules of the form
$\default{R}{[R]}{[R]}$ may be applicable, with 
$\widetilde{R} = \widetilde{P}$ and
$R\subseteq  \widetilde{P}$ (note the relaxed condition here).
This shows $\phi (P, W, 3)= \delta (P)$, and there is no 
additional applicable default rules from now on. \QED
\end{proof}

\end{document}